\documentclass[amsmath,amssymb,showpacs,aip,jcp,reprint,10pt,floatfix]{revtex4-1}

\usepackage{txfonts,hyperref}
\usepackage{color,soul}
\usepackage[T1]{fontenc}
\usepackage{textcomp}
\usepackage{graphicx}
\usepackage[caption=false]{subfig}
\usepackage{comment}
\usepackage{siunitx}
\usepackage[overload]{empheq}

\newcommand\Tstrut{\rule{0pt}{2.6ex}}         % = `top' strut
\newcommand\Bstrut{\rule[-0.9ex]{0pt}{0pt}}   % = `bottom' strut

\soulregister\cite7
\soulregister\ref7
\soulregister\pageref7

\begin{document}

\title{Fingering instabilities in tissue invasion: an active fluid model}
\author{Micha\l{} Bogdan}
\affiliation{Department of Engineering, University of Cambridge, Cambridge, United Kingdom}
\author{Thierry Savin}
\email{t.savin@eng.cam.ac.uk}
\affiliation{Department of Engineering, University of Cambridge, Cambridge, United Kingdom}
\date{9/13/2018}

\begin{abstract}
	Metastatic tumors often invade healthy neighboring tissues by forming multicellular finger-like protrusions emerging from the cancer mass. To understand the mechanical context behind this phenomenon, we here develop a minimalist fluid model of a self-propelled, growing biological tissue. The theory involves only four mechanical parameters and remains analytically trackable in various settings. As an application of the model, we study the evolution of a 2D circular droplet made of our active and expanding fluid, and embedded in a passive non-growing tissue. This system could be used to model the evolution of a carcinoma in an epithelial layer. We find that our description can explain the propensity of tumor tissues to fingering instabilities, as conditioned by both the magnitude of active traction and the growth kinetics. We are also able to derive predictions for the tumor size at the onset of metastasis, and for the number of subsequent invasive fingers. Our active fluid model may help describe a wider range of biological processes, including wound healing and developmental patterning.
\end{abstract}

\maketitle

%%%%%%%%%%%%%%%%%%%%%%%%%%%%%%%%%%%%%%%
%%%%%%%%%%%%%%%%%%%%%%%%%%%%%%%%%%%%%%%
\section{Introduction}\label{sec:introduction}
%%%%%%%%%%%%%%%%%%%%%%%%%%%%%%%%%%%%%%%
%%%%%%%%%%%%%%%%%%%%%%%%%%%%%%%%%%%%%%%

Spreading tumors often do not maintain a straight front while expanding. They instead display an interface patterned with multicellular protrusions, which are commonly referred to as fingers \cite{Tracqui:2009er,Friedl:2009kz}, invading the surrounding tissue \cite{Friedl:2003if,Friedl:2009kz,Tracqui:2009er,Cheung:2013ib,Wong:2014jm,Westcott:2015eb}. Their formation generally initiates cancer metastasis \cite{Friedl:2003if,Friedl:2009kz,Carey:2013gl,Cheung:2013ib,Haeger:2015bh}, which is responsible for the vast majority of cancer-related deaths \cite{Sporn:1996gw}. Similar structures form during wound healing, where fingers accompany re-epithelization \cite{Petitjean:2010hf,Klarlund:2012kg}. In the case of glioblastoma brain tumors, these fingers usually consist of disconnected, diffusing cells \cite{Khain:2006hx,Tracqui:2009er}. But in carcinomas and epithelial wound healing, they tend to remain condensed, with a well-defined boundary \cite{Wong:2014jm,Petitjean:2010hf}.

What causes the formation of such fingers? Studying cancer has traditionally focused on a large number of biological (especially genetic and biochemical) cues \cite{Bizzarri:2008dh}. Yet, these essentially operate by collectively affecting a smaller number of physical properties of the tissues and environments involved \cite{Tracqui:2009er}. How these physical alterations can, in turn, lead to fingering has been investigated by several models. Various causal mechanisms, differing in their assumptions on the mechanical properties adopted for the tissues, have been proposed \cite{Basan:2011bm,Ciarletta:2013em,Goriely:2005iq,BenAmar:2010ig,Basan:2013hb}. For example, reaction-diffusion models of nutrient-limited growth have been used \cite{Tracqui:2009er,Ferreira:2002bx} since the accessibility of diffusing chemicals is necessary for tumor growth. But other models have treated the fingers' emergence as a consequence of a mechanical instability \cite{Guiot:2006ff,BenAmar:2010ig,Ciarletta:2013em,Basan:2011bm}. Support for the latter approach was provided by experiments conducted so as to prevent biochemical signaling, but in which fingering occurred nevertheless \cite{Poujade:2007hm}. Mechanical processes proposed by these models include: fracturing of an elastic surrounding medium (extracellular matrix, healthy tissue) upon pressure from a growing solid inclusion (tumor), leading to subsequent infiltration of the cracks by malignant cells \cite{Guiot:2006ff}; mechanical frustrations between an outer, proliferating ring of a growing tumor and its necrotic core \cite{Ciarletta:2013em}; buckling due to swelling of a spatially restricted gel-like tissue \cite{BenAmar:2010ig}; instability resulting from the interplay between spatially non-uniform cell division/death rates and shear in viscous tissues \cite{Basan:2011bm}; or a pulling mechanism by a subpopulation of leader cells at the tumor's edges \cite{Petitjean:2010hf,Hallou:2017bj}.

Many of the above mechanisms have been successfully described using continuum, analytically solvable models \cite{Basan:2011bm,Ciarletta:2013em,Goriely:2005iq,BenAmar:2010ig}, thus providing deep insights into the physical context involved in tissue fingering. However, these continuum models of tumors have, in most cases, omitted one fundamental component of live tissues: cells actively apply forces (via a conversion of chemical energy into mechanical energy) to their surroundings \cite{Gordon:2003in,duRoure:2005fd,Mierke:2008dg,Trepat:2009bv,Petitjean:2010hf,BlanchMercader:2017fy}. Yet, the onset of fingering at tissue boundaries is often correlated with a dense presence of these so-called active forces and the occurrence of the resulting self-propelled motion \cite{Trepat:2009bv,Petitjean:2010hf}. In discrete models, the central role of these forces in triggering tissue fingering has been verified by simulations within several different frameworks \cite{Hallou:2017bj,Basan:2013hb}. But, to the best of our knowledge, only few continuum models of fingering have studied the effects of activity: the effect of an active rim at the boundary of the tissue has been studied by \citet{Mark:2010dr} and by \citet{Nagilla:2018cx}, while the role of active forces in the tissue bulk has been discussed in wound healing models by \citet{Zimmermann:2014gk} and by \citet{Nesbitt:2017dp}. However, the analyses in the two latter works are limited to a non-dividing, rectangular tissue, either in a static state, or somehow pushed on one side by a rigid barrier.

We here investigate the role of tissue bulk activity for the emergence of fingers in more general situations. We first construct a continuum mechanical model of an active and growing tissue, supported by experimental evidence, and that is analytically solvable and that involves only four physical parameters: friction, activity, growth and surface tension. We next investigate the role of activity in promoting fingering. Provided with experimentally derived estimates of the physical parameters, our model notably produces realistic predictions for the number and evolution of the fingers.

%%%%%%%%%%%%%%%%%%%%%%%%%%%%%%%%%%%%%%%
%%%%%%%%%%%%%%%%%%%%%%%%%%%%%%%%%%%%%%%
\section{Model}\label{sec:model}
%%%%%%%%%%%%%%%%%%%%%%%%%%%%%%%%%%%%%%%
%%%%%%%%%%%%%%%%%%%%%%%%%%%%%%%%%%%%%%%

%%%%%%
%%%%%%
\begin{figure}[ht]
	\includegraphics[width=3.4in]{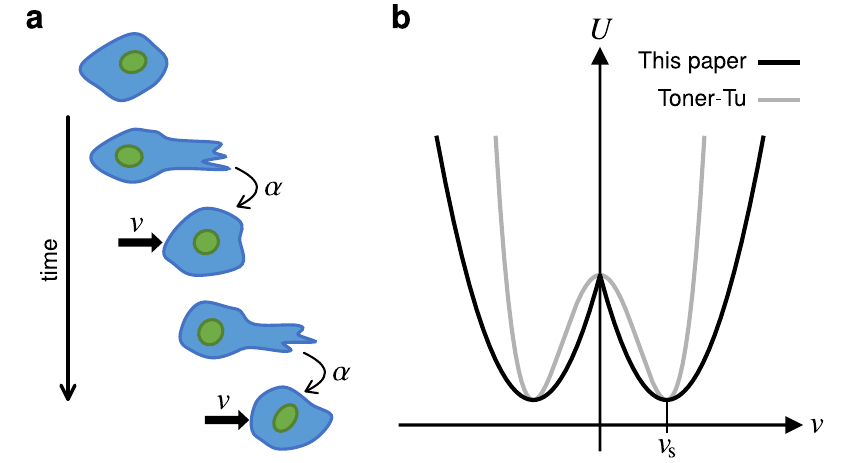}
	\caption{Assumptions of the active traction model. Panel (a) is a schematic illustrating how the active traction of mesenchymal tumor cell acts in the direction of its migration velocity; filopodia and/or lamellipodia are protruding at the leading edge of the cell, which translocates in the same direction. Panel (b) compares (in 1D) the Landau-type ``velocity potential'' \cite{Wensink:2012hk}, $U({\bf v})$ such that the volumic traction force ${\bf F}=-\partial_{\bf v}U$, of our description with the Toner-Tu model; both potentials select an intrinsic velocity $v_\text{s}$ (see text), but our model introduces a discontinuity at ${\bf v}={\bf 0}$ which has no effect on our results. \label{fig1}}
\end{figure}
%%%%%%
%%%%%%

%%%%%%%%%%%%%%%%%%%%%%%%%%%%%%%%%%%%%%%
\subsection{Assumptions and equations}\label{sec:model-assumptions-and-equations}
%%%%%%%%%%%%%%%%%%%%%%%%%%%%%%%%%%%%%%%

In our model of an active and growing tissue, the evolutions of the pressure $p({\bf r},t)$ and velocity ${\bf v}({\bf r},t)$ fields at a position ${\bf r}$ and time $t$ are governed by the following force balance and mass conservation equations,%
%%%%%%
\begin{subequations}\label{eq:evolutions}\begin{align}
		\nabla p             & =-\beta {\bf v}+\alpha\frac{{\bf v}}{|{\bf v}|}\,,\label{eq:force-balance} \\
		\nabla \cdot {\bf v} & =k\,,\label{eq:mass-conservation}
	\end{align}\end{subequations}
%%%%%%
respectively, with $\nabla$ the nabla operator, and where $\alpha$ and $\beta$ are positive parameters, specifying the strength of the interaction between the tissue and a substrate: $\alpha$ describes the magnitude of the active traction, while $\beta$ the magnitude of the effective passive friction (proportional to the tissue viscosity). In Eq.~(\ref{eq:mass-conservation}), $k$ is the net rate of growth (we are here interested in regimes in which it is also positive) of an incompressible tissue, undergoing cell division (or individual cell growth). The $\alpha$-term in Eq.~(\ref{eq:force-balance}), which accounts for the tissue activity, is discussed in detail in Sec.~\ref{sec:activity}. The evaluation of the various parameters is examined in Sec.~\ref{sec:estimation-of-physical-parameters}.

Ignoring the $\alpha$-term, Eq.~(\ref{eq:force-balance}) reduces to Darcy's law $\nabla p =-\beta {\bf v}$ (originally used to describe viscous flows in porous materials and Hele-Shaw apparatus \cite{Saffman:1958cu}), which has been widely employed to model the passive behavior of tissues \cite{Cristini:2003ja,Arciero:2011hj,Ravasio:2015bm,Tracqui:2009er}. Darcy's law notably assumes both a \emph{viscous} and \emph{quasi-2D} dynamics for the deformations of a tissue layer, by considering that the effects of friction against a substrate are much stronger than those of viscous shear within the plane of the layer.

Using two-dimensional models is experimentally justified by the large prevalence of \emph{in vitro} tissue culture monolayers, but also because many \emph{in vivo} soft tissues, including epithelium in which carcinomas develop, tend to spontaneously form quasi-2D monolayers \cite{Trepat:2009bv,Petitjean:2010hf,Friedl:2010ip}. Consequently, two-dimensional descriptions are often employed in tissue mechanics models \cite{Hallou:2017bj,Basan:2011bm,BenAmar:2010ig}.

The mechanical properties of live tissues at short time scales, up to the order of minutes, are generally dominated by an elastic constitutive behavior. At longer timescales, however, a viscous description is better suited \cite{BlanchMercader:2017fy}. The crossover between the two regimes is likely related to the turnover rates of intercellular adherent junctions \cite{Lambert:2007ke,Jennings:2014ut}. Hence, epithelial tissues become fluidized by a reduction in the number of adherent junctions, and a concomitant increase in the magnitude of active traction when becoming malignant. This well-known ``melting'' process is often referred to as the ``epithelial to mesenchymal transition'' \cite{Wong:2014jm,Diepenbruck:2016dr,Friedl:2003if}. Since we here model the behavior of the tissue at timescales on which it experiences substantial growth (that is, on the order of several hours at least \cite{Cheung:2013ib}), the viscous constitutive behavior implied by Darcy's law is justified. Many existing continuum models of epithelial tissues indeed make the same assumption \cite{Arciero:2011hj,Basan:2011bm,BlanchMercader:2017fy,BlanchMercader:2017ft}. Note that, by writing Eqs.~(\ref{eq:evolutions}), we further assume that inertial terms are negligible on these time scales.

%%%%%%%%%%%%%%%%%%%%%%%%%%%%%%%%%%%%%%%
\subsection{Tissue activity}\label{sec:activity}
%%%%%%%%%%%%%%%%%%%%%%%%%%%%%%%%%%%%%%%

The second term on the right-hand side of Eq.~(\ref{eq:force-balance}) accounts for cells actively propelling themselves by exerting traction against the substrate. It will subsequently be referred to as the active term, and $\alpha$ specifies its strength.

We consider here that the direction of the net local active force acting on the tissue layer from the substrate is aligned with the direction of the local flow velocity. This assumption was made in previous studies modeling active tissues \cite{Basan:2013hb,Szabo:2010eu}. It is a consequence of cells attempting to maintain their direction of motility, as illustrated in Fig.~\ref{fig1}a, and also manifested by the persistent Brownian motion of individual cells in vitro \cite{Selmeczi:2005kp}. On a subcellular level, it likely results from the friction destabilizing lamellipodia \cite{Farooqui:2004gg,Klarlund:2012kg} that are not aligned with the cell's velocity \cite{Basan:2013hb}. It has further been shown to fit experimental data \cite{Basan:2013hb}, although this directionality is not universal \cite{Notbohm:2016ew} and other models have been proposed where the direction of the active force to be an independent internal variable, coupled to both stress and velocity fields \cite{Banerjee:2015bq,BlanchMercader:2017ft}.

Eq.~(\ref{eq:force-balance}) also assumes that the active traction does not depend on the magnitude of the velocity. This assumption has been made in several numerical models of motile cells \cite{Basan:2013hb,Mehes:2014jg,Bi:2016bs}, and enables a distinct analysis of the role played by activity.

In the classic theory derived by \citet{Toner:1995jh,Toner:1998ke}, often used to model active fluids \cite{Nesbitt:2017dp,Zimmermann:2014gk,Yang:2015hs}, the net force per unit volume acting on the active fluid from the substrate, in a spatially uniform flow, follows ${\bf F}_{\!\rm T}({\bf v})=\alpha_{\rm T} {\bf v}-\beta_{\rm T} |{\bf v}|^2 {\bf v}$ (ignoring here additional inertial and gradient terms; with $\alpha_{\rm T}$ and $\beta_{\rm T}$ positive parameters). In comparison, our model, Eq.~(\ref{eq:force-balance}), gives this force the expression ${\bf F}({\bf v})=\alpha {\bf v}/|{\bf v}|-\beta {\bf v}$. In both cases, the fluid has a ``preferred'' spontaneous magnitude of velocity $|\mathbf{v}_{\rm s}|=\alpha/\beta$ ($=(\alpha_{\rm T}/\beta_{\rm T})^{1/2}$ in the Toner-Tu model), which it would select when moving in unbounded space without being driven by an external pressure gradient or growth. For magnitudes of velocity lower than $|\mathbf{v}_{\rm s}|$, the fluid would be driven to move faster by the $\alpha$-term, while above it, it would be slowed down by the friction (the $\beta$-term). This fluid's constitutive behavior may be described in terms of a Landau-type ``velocity potential'' \cite{Wensink:2012hk}, shown in Fig.~\ref{fig1}b. Although our model has a discontinuity in the direction of ${\bf F}({\bf v})$ at ${\bf v}=0$ which does not exist in ${\bf F}_{\!\rm T}({\bf v})$, we have verified that this singularity does not significantly affect our subsequent results.

The Toner-Tu model, however, assumes that the friction force grows as $|\mathbf{v}|^3$ and that the active force varies linearly with $|\mathbf{v}|$. Both assumptions are unrealistic when describing biological tissues. Our model, on the other hand, retains the physical interpretation of $\beta$ as the friction coefficient of Darcy's Law, and of $\alpha$ as the magnitude of the active force of the tissue against the substrate (per unit volume). Our approach also enables a direct comparison with classical results for viscous fingering \cite{Paterson:1981df}, readily obtained from our model by taking the limit $\alpha \to 0$.

%%%%%%%%%%%%%%%%%%%%%%%%%%%%%%%%%%%%%%%
\subsection{Estimation of parameters}\label{sec:estimation-of-physical-parameters}
%%%%%%%%%%%%%%%%%%%%%%%%%%%%%%%%%%%%%%%

Based on \emph{in vivo} microscopy observations \cite{Cheung:2013ib} of the time necessary for doubling a carcinoma's size, which is on the order of a few hours, we estimate that the growth rate $k$ is about $10^{-4}\,\text{s}^{-1}$.

The passive friction $\beta$ can be estimated based on \emph{in vitro} force measurements of epithelial tissues against substrates \cite{Pompe:2011ek,Trepat:2009bv,BlanchMercader:2017fy} (admittedly, inferences about \emph{in vivo} systems from these \emph{in vitro} experiments is arguable). Following \citet{Pompe:2011ek}, we assume that friction with the substrate is primarily the consequence of cell-substrate ligands, numbering $200-300$ per cell, each of which exerting a force of about $10^{-12}\,\text{N}$. We thus estimate the total friction force per cellular volume to be about $10^{6}\,\text{N}\,\text{m}^{-3}$ for a $10^{-5}\,\text{m}$ cell size. From \emph{in vivo} microscopy of micrometastasis growth \cite{Cheung:2013ib}, the typical velocity $v$ of the cells falls within $10^{-10}-10^{-9}\,\text{m}\,\text{s}^{-1}$. Hence, dividing the volumic friction force by this velocity provides an estimate $\beta\sim 10^{15}-10^{16}\,\text{Pa}\,\text{s}\,\text{m}^{-2}$.

%%%%%%
%%%%%%
\begin{table}
	\begin{ruledtabular}
		\begin{tabular}{lccr}
			Parameter        & Symbol   & Unit                                 & Value					\Bstrut       \\ \hline
			Growth rate      & $k$      & $\text{s}^{-1}$                      & $10^{-4}$				\Tstrut   \\
			Passive friction & $\beta$  & $\text{Pa}\,\text{s}\,\text{m}^{-2}$ & $10^{15}-10^{16}$  \\
			Active traction  & $\alpha$ & $\text{Pa}\,\text{m}^{-1}$           & $0-10^{10}$        \\
			Surface tension  & $\gamma$ & $\text{Pa}\,\text{m}$                & $10^{-3}- 10^{-2}$ \\
		\end{tabular}
		\caption{Estimates of the physical parameters}
		\label{tab:estimates}
	\end{ruledtabular}
\end{table}
%%%%%%
%%%%%%

There is no lower limit on $\alpha$, as epithelial cells may not exert any active force against the substrate. The upper limit can be estimated on the basis of force tracking microscopy applied to spreading epithelial monolayers \emph{in vitro} \cite{Trepat:2009bv,BlanchMercader:2017fy}. It is observed that traction forces are actively exerted through the monolayers and peak at their edges, giving rise to a gradient of the stress tensor's diagonal terms, which is up to $10^7\,\text{Pa}\,\text{m}^{-1}$ in the study by \citet{Trepat:2009bv}, and $10^8 \,\text{Pa}\,\text{m}^{-1}$ in the work of \citet{BlanchMercader:2017fy}. Balancing $\alpha$ with this typical stress gradient, one can place an upper estimate on $\alpha$ at $\sim10^8 \,\text{Pa}\,\text{m}^{-1}$. Concurring, the traction exerted by single fibroblasts has been reported as up to $10^{-7}-10^{-5}\,\text{N}$ per cell \cite{Ananthakrishnan:2007ba}, which would correspond to $\alpha\sim10^8-10^{10} \,\text{Pa}\,\text{m}^{-1}$ when dividing by the cell's volume.

The effective surface tension of the tissue $\gamma$ will also play a role in our further considerations. Its magnitude depends on the strength of intercellular adhesion and behavior of cortical actin networks \cite{Manning:2010ce,Lecuit:2007cw}. The surface tension was evaluated indirectly in \citet{Foty:1994ba} by measuring the energetic penalty of compression of embryonic multicellular spheroids, revealing values on the order of $3-9\,\text{mPa}\,\text{m}$. We thus presume that $\gamma \sim 10^{-3}- 10^{-2} \,\text{Pa}\,\text{m}$ is a realistic range for our system.

A summary of the estimates for the physical parameters is presented in Tab.~\ref{tab:estimates}.

%%%%%%%%%%%%%%%%%%%%%%%%%%%%%%%%%%%%%%%
\subsection{Model system}\label{sec:system-set-up}
%%%%%%%%%%%%%%%%%%%%%%%%%%%%%%%%%%%%%%%

%%%%%%
%%%%%%
\begin{figure}
	\includegraphics[width=3.4in]{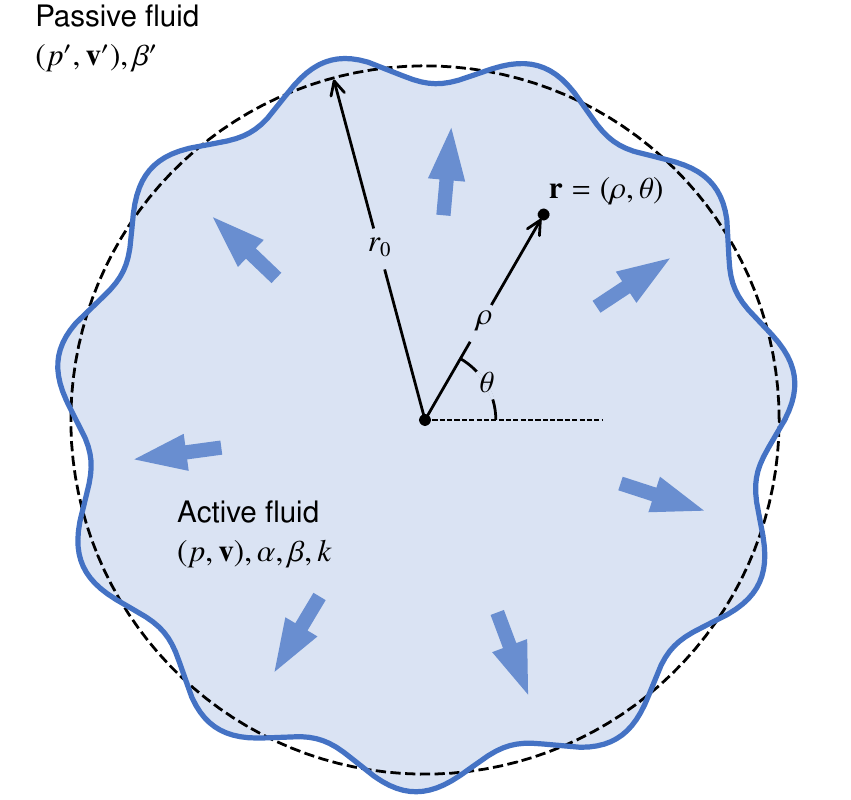}
	\caption{Model system of a tumor growing in an external tissue: a 2D circular droplet with radius $r_0$ and made of an active fluid described by Eqs.~(\ref{eq:evolutions}), is expanding in a passive fluid modeled by Eqs.~(\ref{eq:passive-evolutions}). The interface undergoes periodic perturbations whose linear stability is investigated in Sec.~\ref{sec:results-and-discussion}.\label{fig2}}
\end{figure}
%%%%%%
%%%%%%

The tumor is modeled as an initially circular, 2D droplet of a growing active fluid described by Eqs.~(\ref{eq:evolutions}), with an unperturbed, time-dependent radius $r_0(t)$ (see Fig. \ref{fig2}). The surrounding healthy tissue is modeled as a passive, non-dividing fluid, whose pressure $p'({\bf r},t)$ and velocity ${\bf v}'({\bf r},t)$ fields follow:
%%%%%%
\begin{subequations}\label{eq:passive-evolutions}\begin{align}
		\nabla p'             & =-\beta' {\bf v}'\,,\label{eq:passive-force-balance} \\
		\nabla \cdot {\bf v}' & =0\,,\label{eq:passive-mass-conservation}
	\end{align}\end{subequations}
%%%%%%
where $\beta'$ is the friction parameter (analogous to $\beta$ in the active fluid). In writing Eq.~(\ref{eq:passive-mass-conservation}), we effectively assume that growth in the passive fluid can be neglected on the timescale of metastasis initiation.

We assume that the activity $\alpha$ and the growth rate $k$ are constant (independent of ${\bf r}$ and $t$) through the active tissue.  Constant magnitude of the active force has been assumed in models of active matter before \cite{Marchetti:2016iy,CallanJones:2008jg}. While not correct in all situations\cite{BlanchMercader:2017fy,Kim:2013ex}, it is a convenient assumption to evaluate the influence of its magnitude in fingering. Extensions to non-uniform and time-dependent behaviors of these parameters are readily possible, and we investigate a case of evolving growth rate in Sec.~\ref{sec:selection} (see also appendix~\ref{app:scal-modes-lin}). Similarly, we assume that $\beta$ and $\beta'$ are uniform within their respective regions.

We study the system in polar coordinates ${\bf r}=(\rho,\theta)$ and write vector fields' components in this system with appropriate subscripts, such as ${\bf v}=(v_\rho,v_\theta)$. The perturbed interface between the active and passive fluids, described by the line $r(t,\theta)$, must satisfy two boundary conditions. First, the continuity of the radial components of velocities is expressed as:
%%%%%%
\begin{equation}\label{eq:velocity-bc}
	v_\rho|_{\rho=r} =v'_\rho|_{\rho=r} =\partial_t r\,.
\end{equation}
%%%%%%
Second, the pressure difference across the interface separating the two tissues must equal the Laplace pressure:
%%%%%%
\begin{equation}\label{eq:pressure-bc}
	p|_{\rho=r} - p'|_{\rho=r}=-\gamma \frac{r^2+2(\partial_{\theta}r)^2-r \partial^2_{\theta \theta}r}{[r^2+(\partial_\theta r)^2]^{3/2}}\,,
\end{equation}
%%%%%%
with $\gamma$ the surface tension, and the fraction being the expression of the local interfacial curvature in polar coordinates \cite{doCarmo:1976um}.

Unless stated otherwise, we use the dimensionless variables defined as follows. Distances are rescaled by the characteristic length $\ell=\bigl(\frac{2 \gamma}{\beta k}\bigr)^{1/3}$, which can be interpreted as a capillary length at which growth balances interfacial tension (on the order of $10\,\text{\textmu m}$, based on the estimates of Tab.~\ref{tab:estimates}). Times are rescaled by $k^{-1}$, and we further define $\phi=\beta'/\beta$ the relative viscosity of the displaced tissue compared to the active growing droplet. We introduce a reference activity $\alpha^*=\beta\ell k\sim10^7 \,\text{Pa}\,\text{m}^{-1}$ to make the active traction $\alpha$ dimensionless, $\alpha/\alpha^*\rightarrow\alpha$. Pressures and velocities are made dimensionless by $p^*\equiv \beta \ell^2 k\sim10^2\,\text{Pa}$ and $v^*\equiv \ell k\sim 10^{-9}\,\text{m}\,\text{s}^{-1}$, respectively. We will use the same letters for the dimensionless versions of the variables as for their dimensional counterparts.

The droplet of the active tissue grows due to a positive $k$, as required by Eq.~(\ref{eq:mass-conservation}), and the passive fluid is displaced by it. As long as the interface between the two tissues remains circular (with the unperturbed radius $r_0$), hydrodynamic fields in both regions remain symmetric under rotations and are given by:
%%%%%%
\begin{subequations}\label{eq:symmetric-fields}
	\begin{align}
		\mathbf{v}_0  & =\biggl(\frac{\rho}{2},0\biggr)\,,       & p_0  & =\alpha(\rho-r_0)-\frac{1}{4}(\rho^2-r_0^2)+p|_{\rho=r_0}\,,\label{eq:symmetric-active}                               \\
		\mathbf{v}'_0 & =\biggl(\frac{r_0^2}{2 \rho},0\biggr)\,, & p_0' & =-\frac{\phi r_0^2}{2} \ln\biggl(\frac{\rho}{r_0}\biggr)-\frac{1}{2 r_0}+p|_{\rho=r_0}\,,\label{eq:symmetric-passive}
	\end{align}
\end{subequations}
%%%%%%
as obtained by solving Eqs.~(\ref{eq:evolutions}-\ref{eq:pressure-bc}) and using the dimensionless quantities defined above.

%%%%%%%%%%%%%%%%%%%%%%%%%%%%%%%%%%%%%%%
%%%%%%%%%%%%%%%%%%%%%%%%%%%%%%%%%%%%%%%
\section{Results and discussion}\label{sec:results-and-discussion}
%%%%%%%%%%%%%%%%%%%%%%%%%%%%%%%%%%%%%%%
%%%%%%%%%%%%%%%%%%%%%%%%%%%%%%%%%%%%%%%

%%%%%%%%%%%%%%%%%%%%%%%%%%%%%%%%%%%%%%%
\subsection{Linear stability analysis}\label{sec:prerequisites-of-fingering}
%%%%%%%%%%%%%%%%%%%%%%%%%%%%%%%%%%%%%%%

%%%%%%
%%%%%%
\begin{figure}
	\includegraphics[width=3.4in]{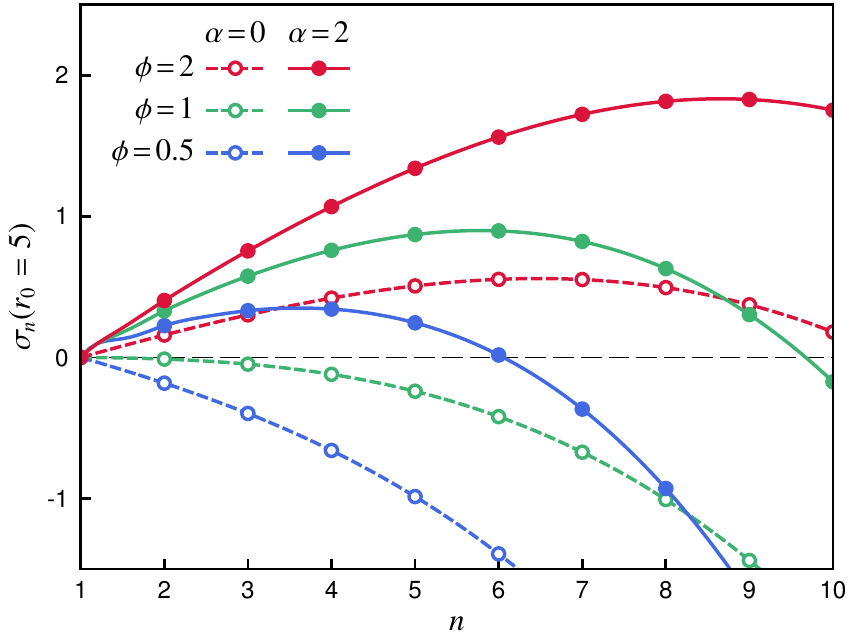}
	\caption{Growth rate $\sigma_{\!n}(r_0=5)$ of periodic interfacial perturbations as a function of the number of fingers $n$, for a droplet of size $r_0=5$, made of a passive ($\alpha=0$; dashed lines) or an active ($\alpha=2$; solid lines) fluid, with varying $\phi$. \label{fig3}}
\end{figure}
%%%%%%
%%%%%%

We investigate under which conditions the active, circular droplet of radius $r_0$ would start to form finger-like protrusions at its edge, while undergoing uniform growth. To do this, we perform a linear stability analysis around the circular solution given in Eqs.~(\ref{eq:symmetric-fields}), by investigating infinitesimal interfacial perturbations of the form $r=r_0+\delta r$, with
%%%%%%
\begin{equation}\label{eq:periodic-ansatz}
	\delta r \propto f_n(t) e^{i n \theta}
\end{equation}
%%%%%%
for an integer $n$ corresponding to the mode of the periodic perturbations, and where $f_n(t)$ is a function describing its time evolution (with $\lim_{t\to0} f_n(t)=1$ for all $n$). An analogous ansatz of periodicity in $\theta$ is made for the perturbations of hydrodynamic fields in both fluids, $(\delta p,\delta p',\delta{\bf v},\delta{\bf v}')\propto f_n(t) e^{i n \theta}$, around the solution given by Eqs.~(\ref{eq:symmetric-fields}). Note that our stability analysis concerns small perturbations around solutions that are themselves time-dependent. Applying the evolutions equations, Eqs.~(\ref{eq:evolutions}-\ref{eq:pressure-bc}), to these perturbed fields provides an expression for the $n$-mode's rate of growth defined by
%%%%%%
\begin{equation}\label{eq:rate-of-growth}
	\sigma_{\!n}(r_0)= \lim_{t \to 0}\frac{\partial_t f_n(t)}{f_n(t)}\,.
\end{equation}
%%%%%%
Positive values of $\sigma_{\!n}(r_0)$ correspond to unstable, growing modes $n$, which may become the basis for the formation of fingers. We derive in appendix~\ref{app:lin-stab} the following expression for $\sigma_{\!n}(r_0)$:
%%%%%%
\begin{equation}\label{eq:active-dispersion-relation}
	\sigma_{\!n}(r_0) =\frac{1}{2}\frac{(\phi-1)(n-1)-n(n^2-1)/r_0^3+\Lambda_n\bigl(\frac{2\alpha}{r_0}\bigr)}{\phi+1+\Lambda_n\bigl(\frac{2 \alpha}{r_0}\bigr)-n\frac{2\alpha}{r_0}}\,,
\end{equation}
%%%%%%
where $\Lambda_n(x)$ is the function
%%%%%%
\begin{equation}\label{eq:lambda-function}
	\Lambda_n(x)= nx-1+n\frac{\sum^n_{k=0} \frac{(-1)^{j}}{n+j}\binom{n+j}{j}\binom{n}{j}x^{-j}}{\sum^n_{j=0} j\frac{(-1)^{j}}{n+j}\binom{n+j}{j}\binom{n}{j}x^{-j}}\,,
\end{equation}
%%%%%%
with $\binom{n}{j}$ the binomial coefficient ``$n$ choose $j$''. Eq.~(\ref{eq:active-dispersion-relation}) holds provided $\alpha<r_0/2$ (see Sec.~\ref{sec:high-activity-regime} for a discussion). The first term in the numerator of Eq.~(\ref{eq:active-dispersion-relation}) represents the effects of the viscosity mismatch, the second term embodies the effects of surface tension, while the final term shows the effects of activity.

Since $\Lambda_n(0)=0$ for all $n$, we obtain the following expression of $\sigma_{\!n}(r_0)$ in the passive limit $\alpha \to 0$:
%%%%%%
\begin{equation}\label{eq:passive-dispersion-relation}
	\sigma_{\!n}(r_0)|_{\alpha=0}=\frac{1}{2}\biggl[\frac{n(\phi-1)}{\phi+1}-1-\frac{n(n^2-1)}{r_0^3(\phi+1)}\biggr]+\frac{1}{\phi+1}\,.
\end{equation}
%%%%%%
The first term (square brackets) of Eq.~(\ref{eq:passive-dispersion-relation}) is equivalent to the landmark result obtained by \citet{Paterson:1981df} for viscous fingering in a radial geometry (Eq.~(10) in Ref.~\onlinecite{Paterson:1981df}), upon imposing the injection rate $Q$ in Paterson's formula equal to the total amount of the droplet's growth per unit time in our setting (that is, $Q=\pi r_0^2$ in dimensionless variables). The last term $\frac{1}{\phi+1}$, however, distinguishes our result from Paterson's, and stems from the fact that, here, the invading fluid also grows within the fingers.

Figure~\ref{fig3} shows $\sigma_{\!n}(r_0)$ vs. $n$ for various values of $\phi$ and $\alpha$, and for an unperturbed droplet radius $r_0=5$. Only integer values of $n$ (circles in Fig.~\ref{fig3}) have a physical interpretation. The first mode $n=1$ corresponds to a translation of the droplet, and since $\Lambda_1(x)= 0$ for all $x$, $\sigma_1(r_0)=0$ for all values of $r_0$, $\alpha$ and $\phi$. Higher modes $n\ge2$ correspond to the formation of $n$ fingers on the interface of the active droplet and, if unstable (that is, if $\sigma_{\!n}(r_0)>0$), could potentially initiate the multicellular protrusions observed in tumors \cite{Cheung:2013ib}.

In passive fluids, $\phi>1$ (that is, the invaded fluid is more viscous than the invading one) is a necessary condition for fingering to be initiated, since instabilities can only grow when the  pressure gradient near the interface is lower in the invading fluid \cite{Flament:1998ga}. However, we observe that $\sigma_{\!n}(r_0)$ increases with $\alpha$, so that modes that are stable when $\alpha=0$ may become unstable in the presence of activity $\alpha>0$ (compare the dashed and solid purple curves, obtained with $\phi=0.5$, in Fig.~\ref{fig3}). Therefore, activity can trigger fingering in systems that are stable otherwise, as well as enhance and/or change the dominant modes in droplets that are already unstable.

Activity lowers the pressure gradient of the invading fluid near the interface, hence promoting instabilities. Using the expression of $p_0$ and $p_0'$ given in Eqs.~(\ref{eq:symmetric-fields}), we find that the condition for fingering, $\nabla\!p |_{\rho=r_0} < \nabla\!p' |_{\rho=r_0}$, is equivalent to $\phi+\frac{2\alpha}{r_0}>1$ when the effects of surface tension are negligible.

%%%%%%%%%%%%%%%%%%%%%%%%%%%%%%%%%%%%%%%
\subsection{High-activity regime}\label{sec:high-activity-regime}
%%%%%%%%%%%%%%%%%%%%%%%%%%%%%%%%%%%%%%%

As already mentioned, $|\mathbf{v}_\text{s}|=\alpha/\beta\sim10^{-9}\,\text{m}\,\text{s}^{-1}$ is a characteristic velocity at which the active fluid would move in an unbounded space, under uniform pressure and without growth. If $|\mathbf{v}_\text{s}|$ exceeds the growth-generated velocity at the interface, the active fluid's motion is frustrated and further instabilities occur across its entire area. We call this regime, for which dimensionless $\alpha>r_0/2$, ``high activity''. In this case, the derivation of Eq.~(\ref{eq:active-dispersion-relation}) presented in the appendix, which assumes that perturbations are only arising at the interface and decaying away from it, is not valid. This regime is potentially relevant in the behavior of real epithelial tissues, in which fingering at the boundaries is accompanied by swirls forming across the entire area of the tissue \cite{Poujade:2007hm}.

Hence, Eq.~(\ref{eq:active-dispersion-relation}) is only valid for the ``low activity'' regime ($\alpha<r_0/2$). Yet, even in this regime, the active droplet may feature a region of instabilities near its center $\rho<2\alpha$, where the velocity magnitude $|\mathbf{v}_0|$ (given by Eq.~(\ref{eq:symmetric-active})) is less than $|\mathbf{v}_\text{s}|$. In particular, this situation would have also occurred in the history of the system considered in Fig.~\ref{fig3}. In practice, a separate simulation-based study would be most appropriate to obtain the velocity field throughout the whole active region, and to further examine the high activity regime. In further sections of this paper, we only examine the system's behavior in the low activity regime.

%%%%%%%%%%%%%%%%%%%%%%%%%%%%%%%%%%%%%%%
\subsection{Onset of fingering}\label{sec:onset-of-fingering}
%%%%%%%%%%%%%%%%%%%%%%%%%%%%%%%%%%%%%%%

%%%%%%
%%%%%%
\begin{figure}
	\includegraphics[width=3.4in]{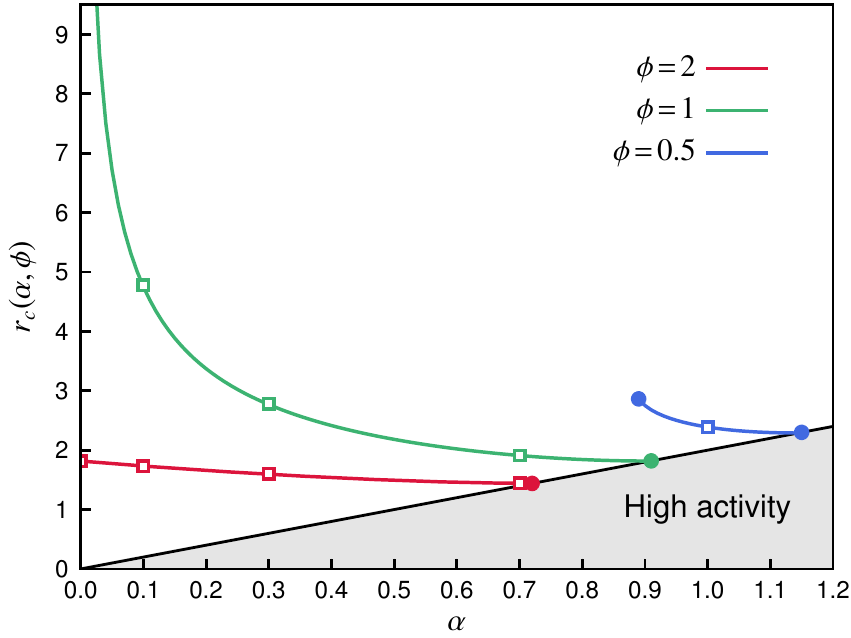}
	\caption{Minimum radius $r_\text{c}(\alpha, \phi)$ for the initiation of fingering. The gray area on the graph signifies the high-activity regime, in which the linear stability analysis does not apply (see sec. \ref{sec:high-activity-regime}). The squares indicate the values of $\alpha$ used in Fig.~\ref{fig5} for each viscosity ratio $\phi$. \label{fig4}}
\end{figure}
%%%%%%
%%%%%%

For small enough radii, surface tension stabilizes the active droplet, but its strength decreases as that droplet grows. Therefore, there exists a critical radius for the onset of fingering, below which the active droplet grows circular and unperturbed. Since $n=2$ is always the first mode to become unstable, $r_\text{c}(\alpha, \phi)$ defined by the conditions $\sigma_2( r_0=r_c)=0$ and $\partial_{r_0}\sigma_2(r_0)|_{r_0=r_\text{c}}>0$ provides an estimate of that critical radius. Using Eq.~(\ref{eq:active-dispersion-relation}) with $\Lambda_2(x)=x(3x-4)/(2x-3)$, these conditions are equivalent to finding a polynomial root (first condition) within a subdomain (second condition), and have a unique positive solution.

We plot $r_\text{c}(\alpha, \phi)$ vs. $\alpha$ in Fig.~\ref{fig4}, which shows that increasing activity decreases the minimum radius for fingering. When $\phi<1$ (purple curve in Fig.~\ref{fig4}), there exists a minimum value of $\alpha$, below which fingering cannot occur, because the higher viscosity of the invading droplet has a stabilizing effect. When $\phi = 1$ (green curve in Fig.~\ref{fig4}), a moderate increase of $\alpha$ may decrease $r_\text{c}$ multiple times.  The impact of $\alpha$ in fingering is, however, reduced when $\phi>1$ (the red curve in Fig.~\ref{fig4}), since in that case the interface would be unstable even without activity.

The range of dimensionless $r_\text{c}$ presented in Fig.~\ref{fig4} would correspond to a radius of $10-100\,\text{\textmu m}$; however, it can be much higher for lower values of $\alpha$. This range of $r_\text{c}$ is nevertheless in qualitative agreement with the tumor size at which the onset of fingering occurred in experimentally studied carcinomas\cite{Cheung:2013ib}.

%%%%%%%%%%%%%%%%%%%%%%%%%%%%%%%%%%%%%%%
\subsection{Dominant mode}\label{sec:selection}
%%%%%%%%%%%%%%%%%%%%%%%%%%%%%%%%%%%%%%%

We now address the question of how many fingers are visible in practice, or, technically speaking, the question of which perturbation mode dominates during growth. Viscous fingering studies suggest the dominant mode is the one satisfying the so-called maximum-amplitude criterion \cite{Dias:2013dx}. The criterion is satisfied by the mode $n_{\text{d}}$ experiencing the largest total aggregated growth in amplitude $\zeta_n$ over the entire history of the system \cite{Dias:2013dx}. Following Ref.~\onlinecite{Dias:2013dx}, we obtain $\zeta _n$ by integrating the rate of the perturbation's growth, as predicted by our linear stability analysis, over that history:
%%%%%%
\begin{equation}\label{eq:max-criterion}
	\zeta _n(r_0)=\exp\biggl[\int^{r_0}_{R_n}\sigma_{\!n}(r)\frac{{\rm d} t}{{\rm d} r} {\rm d} r \biggr]\,,
\end{equation}
%%%%%%
where $R_n$ is the radius at which mode $n$ is first destabilized (i.e., the minimum radius at which $\sigma_{\!n}(R_n)$ becomes positive). The dominant mode $n_{\text{d}}$ is then obtained for each $r_0$ from the conditions,
%%%%%%
\begin{subequations}\label{eq:dominant-mode-conditions}
	\begin{align}
		\partial_n \zeta_n(r_0) |_{n=n_{\text{d}}}      & =0\,,\label{eq:dominant-mode-condition-1} \\
		\partial^2_{nn} \zeta_n(r_0) |_{n=n_{\text{d}}} & <0\,,\label{eq:dominant-mode-condition-2}
	\end{align}
\end{subequations}
%%%%%%
used to locate the maximum aggregated growth.

As we shall see, the selection of the dominant mode depends on the particular kinetics of the tumor growth. Some experimental studies have shown that an initially exponential growth \cite{Withers:2006fd} (corresponding to a constant $k$) subsequently slows down with time (implying a decrease in average $k$) as the tumor enlarges and its resource supply becomes a limiting factor \cite{Herman:2011ey,Guiot:2003hd,Benzekry:2014kc}. Other kinetics have been measured for various tumors and phases of growth, including sigmoidal regimes in which the growth stalls \cite{Herman:2011ey,Guiot:2003hd,Benzekry:2014kc}. A kinetics where the tumor's radius grows linearly with time also naturally emerges when the tumor proliferates only within an outer rim \cite{Benzekry:2014kc}. Such growth can also occur as a temporary feature in a sigmoidal kinetics.

We thus proceed to discuss in detail the selection of dominant modes in two kinetic models of tumor growth: an exponentially growing tumor, where $k$ is uniform and independent of time and $r_0(t)=r_\text{i}e^{k(t-t_\text{i})/2}$ (from Eq.~(\ref{eq:mass-conservation}) at the interface, and with dimensional variables; $r_\text{i}$ being the initial radius at time $t_\text{i}$); and a tumor with a radius growing linearly with time, $r_0(t)=r_\text{i}+\upsilon(t-t_\text{i})$, with $\upsilon$ the constant and uniform velocity of the unperturbed interface. In the latter case, the growth rate $k$ appearing in Eq.~(\ref{eq:mass-conservation}) evolves with time.

\subsubsection{Exponential growth}
%%%%%%%%%%%%%%%%%%%%%%%%%%%%%%%%%%%%%%%

%%%%%%
%%%%%%
\begin{figure}
	\includegraphics[width=3.4in]{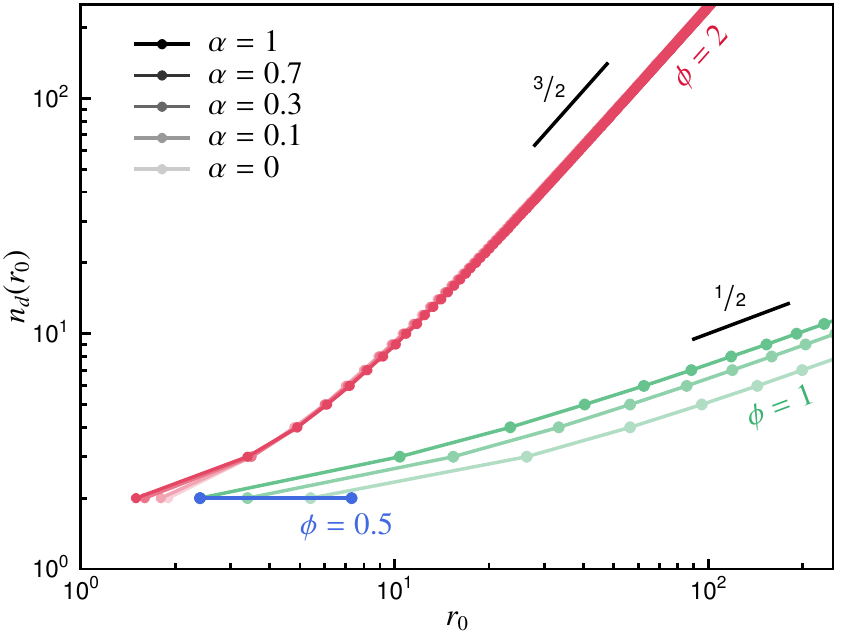}
	\caption{Numerical estimation of the dominant mode $n_\text{d}(r_0)$ observed in an active droplet of radius $r_0$, undergoing exponential growth, for various values of $\alpha$ and $\phi$. \label{fig5}}
\end{figure}
%%%%%%
%%%%%%

The integral given in Eq.~(\ref{eq:max-criterion}) cannot be expressed analytically for all values of $\phi$ and $\alpha$, and we plot in Fig.~\ref{fig5} the relationship $n_{\text{d}}(r_0)$ obtained from numerical evaluation. When $\phi>1$, and for low radii close to the onset of fingering, we observe that the activity has only a moderate influence on the selection of the dominant mode. For later growth, when $r_0$ becomes large, the viscosity mismatch is the governing cause of fingering and the activity plays no role in the selection of the dominant mode. We numerically observe the power-law variations $n_{\text{d}}\propto r_0^{3/2}$, independent of $\alpha$. When $\phi=1$, higher activities promote the selection of higher modes. We obtain numerically, and for large $r_0$, the scaling $n_{\text{d}}\propto r_0^{1/2}$, where the $1/2$ power law is independent of $\alpha$. When $\phi<1$, certain low-$n$ modes will become destabilized, provided sufficient $\alpha$. However, these perturbations will restabilize and decay as $r_0$ increases further, because the stabilization from viscosity mismatch dominates as the radius of the droplet grows: active terms of Eq.~(\ref{eq:active-dispersion-relation}) vanish $r_0 \to \infty$, while terms involving $\phi$ remain constant in this limit.

We may recover analytically the observed scalings for large $n_{\text{d}}$ and $r_0$, and for $\phi \ge 1$. We derive in the appendix~\ref{app:scal-modes-exp} the following results when $r_0\to\infty$: $n_\text{d}\approx (\phi-1)^{1/2}\times r_0^{3/2}$ for $\phi>1$ and $n_\text{d}\approx\bigl[(\alpha/2)^2+(\alpha/2)^{1/2}\bigr]^{1/2}\times r_0^{1/2}$ for $\phi=1$, which we give below in dimensional variables to highlight the influence of the various physical parameters:
%%%%%%
\begin{subequations}\label{eq:scaling_exp}
	\begin{align}[left =n_\text{d}\approx\empheqlbrace\,]
		& c^{1/2} \biggl[\frac{k(\beta'-\beta)}{2\gamma}\biggr]^{1/2}\times r_0^{3/2}                                                                  & \text{for $\beta'>\beta$}\,,\label{eq:scaling-1} \\
		& \biggl[\frac{2\gamma}{\beta k}\biggl(\frac{\alpha}{4\gamma}\biggr)^2+\biggl(\frac{\alpha}{4\gamma}\biggr)^{1/2}\biggr]^{1/2}\times r_0^{1/2} & \text{for $\beta'=\beta$}\,,\label{eq:scaling-2}
	\end{align}
\end{subequations}
%%%%%%
when $r_0\gg \bigl(\frac{2 \gamma}{\beta k}\bigr)^{1/3}$, and with $c \approx 0.06$ defined as the smaller of the two solutions to $3c=3+\ln c$.

\subsubsection{Linear growth}
%%%%%%%%%%%%%%%%%%%%%%%%%%%%%%%%%%%%%%%

We also investigate pattern selection for a tumor with a radius growing linearly with time. The derivation of $\sigma_{\!n}(r_0)$ proceeds along identical lines, although in this case, the integral in Eq.~(\ref{eq:max-criterion}) can be expressed analytically. Details of this calculation are given in the appendix~\ref{app:scal-modes-lin}, and in this case we find that $n_\text{d} \approx c(\alpha+\phi-1)^{1/2}\times r_0$ is valid for all values of the physical parameters in the low activity regime when $r_0\to\infty$, and where $c \approx 0.06$ is the constant defined previously. We thus write, in dimensional form,
%%%%%%
\begin{equation}\label{eq:scaling_lin}
	n_\text{d} \approx c \biggl[\frac{\alpha+\upsilon(\beta'-\beta)}{\gamma}\biggr]^{1/2}\times r_0\,,
\end{equation}
%%%%%%
when $r_0\gg \bigl(\frac{\gamma}{\beta \upsilon}\bigr)^{1/2}$ (note that the dimensionless variables are defined differently in the linear growth, as explained in appendix~\ref{app:scal-modes-lin}). The difference in the $n_\text{d}$ vs. $r_0$ power-law dependency between Eqs.~(\ref{eq:scaling_exp}) and (\ref{eq:scaling_lin}) highlights the role of the growth kinetics in the fingering pattern, and is discussed with more details in the following.

\subsubsection{Comparison and discussion}
%%%%%%%%%%%%%%%%%%%%%%%%%%%%%%%%%%%%%%%

%%%%%%
%%%%%%
\begin{figure}
	\includegraphics[width=3.4in]{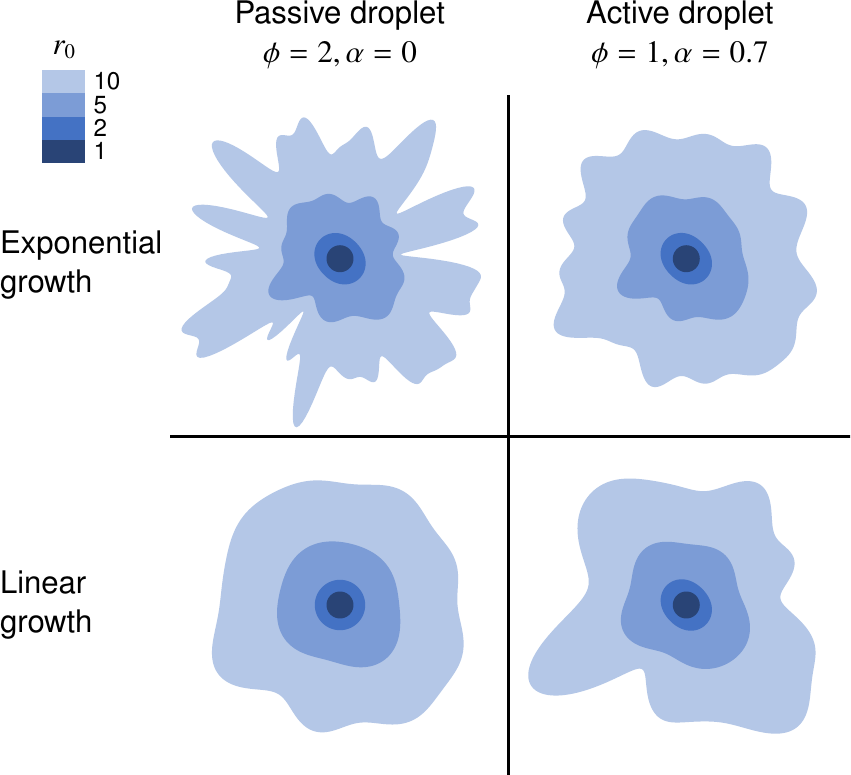}
	\caption{Evolution of a droplet, undergoing exponential (upper row) or linear (lower row) growth, made of a passive (left column) or active (right column) fluid. For comparison, the patterns are shown at the indicated values of the unperturbed radii $r_0$ (see the scale of shades), even though these are reached at different times in each growth regime. We used the initial amplitude $0.2$ for all modes and, to facilitate the morphological comparison, equate the characteristic lengths introduced for each growth kinetics (see $\ell$ defined in Sec.~\ref{sec:system-set-up} for the exponential growth, and in appendix~\ref{app:scal-modes-lin} for the linear growth): $\bigl(\frac{2 \gamma}{\beta k}\bigr)^{1/3}=\bigl(\frac{\gamma}{\beta \upsilon}\bigr)^{1/2}\Leftrightarrow 4\beta\upsilon^3=\gamma k^2$.\label{fig6}}
\end{figure}
%%%%%%
%%%%%%

We now examine the evolution of the tumor's shape in the two growth kinetics studied above. An initially circular droplet is allowed to evolve, with the $n$-mode perturbation starting when the radius $r_0$ reaches $R_n$, and with an initial amplitude of $0.2$ (corresponding to $\sim2\,\text{\textmu m}$). The perturbation is subsequently allowed to grow according to Eq.~(\ref{eq:max-criterion}), such that $\zeta_n(r_0)$ represents the weight of the $n$-mode at the unperturbed droplet radius $r_0$. We further assigned a random phase difference between each $n$-mode perturbation.

We present in Fig.~\ref{fig6} examples of droplet patterns obtained with this procedure, where fingering is driven by either viscosity mismatch (left) or by activity (right), in both the exponential (top) and linear (bottom) growth regimes. In the linear growth, activity-driven fingers emerge more distinctively than in the passive droplet; the opposite is observed in the exponential growth. These results, as well as the analytical scalings presented above, demonstrate that the role of activity in fingering depends on the kinetics of the tumor's growth, and is indeed enhanced in the slower, linear growth kinetics. This assessment could potentially provide a basis for the mechanism behind the onset of metastasis, when the bulk growth of the primary tumor slows down or saturates.

The results presented in Figs.~\ref{fig5} and \ref{fig6} relate direct observables of the tumor's geometry, and such measurements should indeed be envisaged by experimentalists. Note, however, from Fig.~\ref{fig6} that the number of fingers ($\sim10$) visible at $r_0\sim100\,\text{\textmu m}$ when $\phi=1$ is in agreement with the experimental observations shown by \citet{Cheung:2013ib}.

%%%%%%%%%%%%%%%%%%%%%%%%%%%%%%%%%%%%%%%
%%%%%%%%%%%%%%%%%%%%%%%%%%%%%%%%%%%%%%%
\section{Conclusions}
%%%%%%%%%%%%%%%%%%%%%%%%%%%%%%%%%%%%%%%
%%%%%%%%%%%%%%%%%%%%%%%%%%%%%%%%%%%%%%%

We have devised a model of a growing and self-propelled tissue that isolates the role of four mechanical parameters (summarized in Tab.~\ref{tab:estimates}) on its dynamics. The theory is based on experimental evidence and analytically trackable. We used it to describe the evolution of an embedded 2D circular droplet that could model a carcinoma in an epithelial layer. In this example, we were able to highlight the basic mechanical conditioning required to form interfacial instabilities, reminiscent of the classical viscous fingering, and that could explain the tumor protrusions observed at the onset of metastasis. We notably find that the tissue's active traction and growth kinetics are central to shape the instabilities' pattern and evolution.

Our model, and the example of its application presented here, could further help predict the minimum tumor size for metastasis, as well as the number of subsequent invasive fingers emerging from the initial mass. To the best of our knowledge, these observable geometric quantities have yet to be measured systematically in experimental studies.

The relative analytical simplicity of our model allows the investigation of more complex settings, such as heterogeneous tumors where active forces and/or growth are not uniform, or where these parameters are evolving with time. It also offers constitutive equations that can be used in simulations, and we envisage such studies for systems with high traction forces, where active motions are faster than the growth velocity, and which may indeed be relevant in aggressive forms of cancer.

%%%%%%%%%%%%%%%%%%%%%%%%%%%%%%%%%%%%%%%
%%%%%%%%%%%%%%%%%%%%%%%%%%%%%%%%%%%%%%%
\section{Acknowledgements}
%%%%%%%%%%%%%%%%%%%%%%%%%%%%%%%%%%%%%%%
%%%%%%%%%%%%%%%%%%%%%%%%%%%%%%%%%%%%%%%

The authors thank Drs. J. Prost, S. Lira, A. Hallou, and P. Szymczak for insightful discussions.

\appendix
%%%%%%%%%%%%%%%%%%%%%%%%%%%%%%%%%%%%%%%
%%%%%%%%%%%%%%%%%%%%%%%%%%%%%%%%%%%%%%%
\section{Linear stability analysis}\label{app:lin-stab}
%%%%%%%%%%%%%%%%%%%%%%%%%%%%%%%%%%%%%%%
%%%%%%%%%%%%%%%%%%%%%%%%%%%%%%%%%%%%%%%

We here follow the lines of the demonstration given by \citet{Paterson:1981df}. We substitute the linearized perturbations to the hydrodynamic fields in both fluids, $(\delta p,\delta p',\delta{\bf v},\delta{\bf v}')$, into Eqs.~(\ref{eq:evolutions}-\ref{eq:passive-evolutions}) and obtain, to linear order:
%%%%%%
\begin{subequations}\label{eq:act-perturb}\begin{align}
		\partial_{\rho} \delta p             & =-\delta v_{\rho}\,,\label{eq:act-perturb1}                                            \\
		v_{\rho}\partial_{\theta}\delta p    & =\alpha \rho\delta v_{\theta}-\rho v_{\rho}\delta v_{\theta}\,,\label{eq:act-perturb2} \\
		\partial_{\rho}(\rho\delta v_{\rho}) & =-\partial_{\theta} \delta v_{\theta}\,,\label{eq:act-perturb3}
	\end{align}\end{subequations}
%%%%%%
in the active fluid, and
%%%%%%
\begin{subequations}\label{eq:pas-perturb}\begin{align}
		\partial_{\rho} \delta p'             & =-\phi\delta v'_{\rho}\,,\label{eq:pas-perturb1}                 \\
		\partial_{\theta}\delta p             & =-\rho \phi\delta v'_{\theta}\,,\label{eq:pas-perturb2}          \\
		\partial_{\rho}(\rho\delta v'_{\rho}) & =-\partial_{\theta} \delta v'_{\theta}\,,\label{eq:pas-perturb3}
	\end{align}\end{subequations}
%%%%%%
in the passive fluid.

Introducing the ansatz of periodicity in $\theta$, that is $(\delta p,\delta p',\delta{\bf v},\delta{\bf v}')\propto e^{i n \theta}$, in Eqs.~(\ref{eq:act-perturb}-\ref{eq:pas-perturb}) allows us to calculate the partial derivatives with respect to $\theta$. Combining the resulting equations, and using the expression of ${\bf v}_0$ of Eq.~(\ref{eq:symmetric-active}), leads to second-order differential equations for the pressure perturbations in both fluids:
%%%%%%
\begin{subequations}\label{eq:diff-press}\begin{align}
		n^2 \delta p  & =(\rho-2 \alpha)\partial_{\rho}(\rho\partial_{\rho} \delta p)\,,\label{eq:act-diff-press} \\
		n^2 \delta p' & =\rho\partial_{\rho}(\rho\partial_{\rho} \delta p')\,.\label{eq:pas-diff-press}
	\end{align}\end{subequations}
%%%%%%
We now assume $\alpha<r_0/2$ (see Sec.~\ref{sec:high-activity-regime}). Eqs.~(\ref{eq:diff-press}) are both solved by linear combinations of 2 functions. One function in each of these combinations has incorrect asymptotic behavior (diverging, instead of decaying away from the interface) and is removed. The retained functions serve to formulate the allowed forms of pressure perturbations:
%%%%%%
\begin{subequations}\label{eq:press-perturb}\begin{align}
		\delta p  & = q(t) \biggl[\sum^n_{j=0} \frac{(-1)^{j}}{n+j}\binom{n+j}{j}\binom{n}{j}\biggl(\frac{2 \alpha}{\rho}\biggr)^{-j}\biggr]e^{in\theta}\,,\label{eq:act-press-perturb} \\
		\delta p' & = q'(t)\biggl(\frac{\rho}{r_{0}}\biggr)^{-n}e^{in\theta}\,.\label{eq:pas-press-perturb}
	\end{align}\end{subequations}
%%%%%%
The functions $q(t)$ and $q'(t)$ are then obtained by finding $\delta v_{\rho}$ (using Eq.~(\ref{eq:act-press-perturb}) in Eq.~(\ref{eq:act-perturb1})) and $\delta v'_{\rho}$ (using Eq.~(\ref{eq:pas-press-perturb}) in Eq.~(\ref{eq:pas-perturb1})), and substituting the resulting expressions into the kinematic boundary conditions, Eq.~(\ref{eq:velocity-bc}). One gets:
%%%%%%
\begin{subequations}\label{eq:press-final}\begin{align}
		\delta p  & = \frac{r_0}{n}\biggl[\partial_t \delta r-\frac{\delta r}{2}\biggr]\biggl[1+\frac{2 \alpha}{r_0}-\Lambda_n\biggl(\frac{2 \alpha}{\rho}\biggr)-(n\!-\!1)\frac{2 \alpha}{\rho}\biggr]\,,\label{eq:act-press-final} \\
		\delta p' & =\frac{\phi r_0}{n}\biggl[\partial_t \delta r+\frac{\delta r}{2}\biggr]\biggl(\frac{\rho}{r_0}\biggr)^{-n}\,,\label{eq:pas-press-final}
	\end{align}\end{subequations}
%%%%%%
with $\Lambda_n$ the function defined by Eq.~(\ref{eq:lambda-function}), and where the $\theta$-dependency $e^{in\theta}$ has been incorporated into $\delta r$. Using both expressions into the pressure boundary condition, Eq.~(\ref{eq:pressure-bc}) linearized to first order in $\delta r$, and substituting $\delta r\propto f_n(t)e^{in\theta}$ leads to an expression for $\partial_t f_n/f_n$. Upon using the definition of $\sigma_{\!n}(r_0)$, Eq.~(\ref{eq:rate-of-growth}), we finally obtain Eq.~(\ref{eq:active-dispersion-relation}).

%%%%%%%%%%%%%%%%%%%%%%%%%%%%%%%%%%%%%%%
%%%%%%%%%%%%%%%%%%%%%%%%%%%%%%%%%%%%%%%
\section{Scaling laws for the dominant modes}\label{app:scal-modes}
%%%%%%%%%%%%%%%%%%%%%%%%%%%%%%%%%%%%%%%
%%%%%%%%%%%%%%%%%%%%%%%%%%%%%%%%%%%%%%%

%%%%%%%%%%%%%%%%%%%%%%%%%%%%%%%%%%%%%%%
\subsection{Exponential growth}\label{app:scal-modes-exp}
%%%%%%%%%%%%%%%%%%%%%%%%%%%%%%%%%%%%%%%

We obtain the scaling of Eq.~(\ref{eq:scaling-1}) for the viscosity-controlled fingering, $\phi>1$, by only considering the passive case ($\alpha=0$). The result is valid even when $\alpha>0$, since effects of the viscosity mismatch dominate effects of activity when $r_0 \to \infty$, as also indicated by the numerical results shown in Fig.~\ref{fig5}. We then use the expression $\sigma_{\!n}(r_0)|_{\alpha=0}$ given by Eq.~(\ref{eq:passive-dispersion-relation}) into Eq.~(\ref{eq:max-criterion}), in which we substitute ${\rm d}t=2{\rm d}r/r$ for exponential growth. We calculate
%%%%%%
\begin{equation*}
	\zeta _n(r_0)|_{\alpha=0}=\bigl[A_n\exp(A_n^{-1}-1)\bigr]^{B_n}\,,
\end{equation*}
%%%%%%
with $A_n=\frac{\phi-1}{n(n+1)}r_0^3$ and $B_n=\frac{(n-1)(\phi-1)}{3(\phi+1)}$, by using the expression of $R_n$ obtained from solving $\sigma_{\!n}(R_n)|_{\alpha=0}=0$. The dominant mode $n_{\text{d}}$ is calculated for each $r_0$ from the conditions of Eqs.~(\ref{eq:dominant-mode-conditions}). Taking the limit $r_0 \to \infty$, and reintroducing dimensional variables, lead to Eq.~(\ref{eq:scaling-1}).

When $\phi=1$, the selection of the dominant mode depends on the activity $\alpha$. We first Taylor expand $\Lambda_n(x)$ to first order in $x$,
%%%%%%
\begin{equation}\label{eq:approx-lambda}
	\Lambda_n(x) \approx \frac{2n(n-1)}{2n-1}x\,.
\end{equation}
%%%%%%
This term grows linearly with $n$ for $n \to \infty$, while higher order coefficients in $x$ plateau in this limit. Therefore, this expression of $\Lambda_n$ is efficient even for $x\sim1$, and we use it to approximate $R_n \approx \bigl[\frac{(n+1)(2n-1)}{4 \alpha}\bigr]^{1/2}$ and $\sigma_{\!n}(r_0)\approx \frac{n(n-1)}{4r_0^3}\bigl[\frac{\alpha (n^2+n-4 r^3_0)}{(2 n-1) r_0}-n-1\bigr]$. Upon substituting these approximations into Eq.~(\ref{eq:max-criterion}),  the dominant mode conditions Eqs.~(\ref{eq:dominant-mode-conditions}) lead to Eq.~(\ref{eq:scaling-2}) for $r_0 \to \infty$.

%%%%%%%%%%%%%%%%%%%%%%%%%%%%%%%%%%%%%%%
\subsection{Linear growth}\label{app:scal-modes-lin}
%%%%%%%%%%%%%%%%%%%%%%%%%%%%%%%%%%%%%%%

In this kinetics, the growth rate $k$ is not a constant, as opposed to the velocity $\upsilon$ of the unperturbed interface which is now the adequate growth parameter. We thus use a different set of dimensionless variables, based on $\upsilon$. The characteristic length is now $\ell=\bigl(\frac{\gamma}{\beta \upsilon}\bigr)^{1/2}$, and times are rescaled by $\ell/\upsilon$, activities by $\alpha^*=\beta \upsilon$, pressures by $p^*\equiv \beta \ell \upsilon$  and velocities by $\upsilon$. The derivation presented in appendix~\ref{app:lin-stab} proceeds similarly, and we find that the rate of perturbation's growth $\sigma_n(r_0)$ in this system of variables is written as:
%%%%%%
\begin{equation}\label{eq:dispersion-rel-linear-growth}
	\sigma_{\!n}(r_0) =\frac{1}{r_0}\frac{(\phi-1)(n-1)-n(n^2-1)/r_0^2+\Lambda_n\bigl(\alpha\bigr)}{\phi+1+\Lambda_n\bigl(\alpha\bigr)-n\alpha}\,.
\end{equation}
%%%%%%
The integral of Eq.~(\ref{eq:max-criterion}), with the change of variables now written ${\rm d}t={\rm d}r$, may then be calculated analytically and we obtain:
%%%%%%
\begin{equation}
	\zeta_n(r_0)=\bigl[A_n\exp(A_n^{-1}-1)\bigr]^{B_n}
\end{equation}
%%%%%%
with $A_n=\frac{(\phi-1)(n-1)+\Lambda_n(\alpha)}{n(n^2-1)}r^2_0$ and $B_n=\frac{1}{2}\frac{(\phi-1)(n-1)+\Lambda_n(\alpha)}{\phi+1+\Lambda_n(\alpha)-n\alpha}$. Upon using the approximation of $\Lambda_n(x)$ given by Eq.~(\ref{eq:approx-lambda}), the conditions Eqs.~(\ref{eq:dominant-mode-conditions}) now imply Eq.~(\ref{eq:scaling_lin}) in the limit $r_0 \to \infty$.

%%%%%%%%%%%%%%%%%%%%%%%%%%%%%%%%%%%%%%%
\section*{References}
%%%%%%%%%%%%%%%%%%%%%%%%%%%%%%%%%%%%%%%
\bibliography{ArXiv-Savin2018}
\end{document}